# Molecular formation along the atmospheric mass loss of HD 209458 b and similar Hot Jupiters

R. Pinotti [a,b], H.M. Boechat-Roberty [a,*]

[a] Observatório do Valongo, Universidade Federal do Rio de Janeiro - UFRJ, Ladeira Pedro Antônio 43, 20080-090, Rio de Janeiro, RJ, Brasil
[b] Petrobras, Av. Henrique Valadares, 28, A1, 9th floor, 20231-030, Rio de Janeiro, RJ, Brazil

___

**Abstract:**

The chemistry along the mass loss of Hot Jupiters is generally considered to be simple, consisting mainly of atoms, prevented from forming more complex species by the intense radiation field from their host stars. In order to probe the region where the temperature is low (T < 2000 K), we developed a 1D chemical and photochemical reaction model of the atmospheric mass loss of HD 209458 b, involving 56 species, including carbon chain and oxygen bearing ones, interacting through 566 reactions. The simulation results indicate that simple molecules like $OH^+$, $H_2O^+$ and $H_3O^+$ are formed inside the region, considering that residual $H_2$ survives in the exosphere, a possibility indicated by recent observational work. The molecules are formed and destroyed within a radial distance of less than $10^7$ km, but the estimated integrated column density of $OH^+$, a potential tracer of $H_2$, is high enough to allow detection, which, once achieved, would indicate a revision of chemical models of the upper atmosphere of Hot Jupiters. For low density Hot Jupiters receiving less intense XUV radiation from their host stars than HD 209458 b, molecular species could conceivably be formed with a higher total column density.

___

## 1. Introduction

The discovery of extrasolar planets orbiting main sequence stars, which began with the discovery of 51 Peg b in 1995 (Mayor and Queloz, 1995), has uncovered a class of planet not foreseen by theorists, nor observed in our Solar System: the so called Hot Jupiters, gas giant planets orbiting their host stars at distances of 0.1 AU or less. Some of them are so close that their outer atmospheres are continuously losing mass. There are several studies dedicated to modeling the mass loss of Hot Jupiters, particularly of HD 209458 b (Bourrier and Lecavelier des Etangs, 2013; Guo, 2013; 2011; Koskinen et al., 2013a, 2013b; Murray-Clay et al., 2009; Penz et al., 2008; Garcia-Muñoz, 2007). Nevertheless, the study of chemical reactions of molecules along the atmospheric mass loss has been overlooked, supposedly because it is thought that the intense UV radiation field of the host stars would prevent their formation. However, these studies are generally limited to a distance of five times the planet's radius (Garcia Munoz, 2007; Koskinen et al., 2013a, 2013b).

The radial velocity method for finding exoplanets favors a star-planet system with a higher planet mass and/or a smaller orbit (Perryman, 2011). Therefore, it is no surprise, looking in retrospect, which the first exoplanet discovered around a main sequence star was itself a Hot Jupiter, 51 Peg b, with a minimum mass of 0.47 $M_J$ and an orbital semi-major axis of 0.052 AU. The improvement in the sensitivity of the method over time, together with the increasing success of other methods, particularly the transit method, have allowed a less biased statistics of extrasolar planets, where Hot Jupiters play a minor role. Still, Hot Jupiters account for around 20% of all known planets, and are expected to be present in around 1.2% of all F, G and K dwarfs in the Solar vicinity (Wright et al., 2012), although this estimate is still not well established (Wang J. et al., 2015).

In spite of the apparent minor role of Hot Jupiters in the general exoplanet population, their existence has spurred a wealth of research, mainly on theories of planetary formation and migration, since Hot Jupiters are thought to have been formed further out in the protoplanetary disk, with subsequent inward migration (Perryman, 2011; Lubow and Ida, 2010). Studies of atmospheric composition of extrasolar gas giants have also been allowed for transiting Hot Jupiters, through spectroscopy and photometry during primary and secondary transits.

The most studied transiting Hot Jupiter is likely to be HD 209458 b (Linsky et al. 2010), that has a mass of 0.64 $M_J$ and an orbital radius of 0.0475 AU. This exoplanet, also known as Osiris, is orbiting HD 209458, a G0 V type star that is relatively nearby (47 pc). Charbonneau et al. (2002) made the first detection of an extrasolar atmosphere during a transit of HD 209458 b. Many molecules, such as methane ($CH_4$), water vapor ($H_2O$) and carbon dioxide ($CO_2$), have been detected in its atmosphere (Swain et al., 2009).

___

*Corresponding author. Email address: heloisa@astro.ufrj.br

Vidal-Madjar et al. (2003) discovered that the Lyman-α line flux from the star HD 209458 was reduced by ~15% during primary transit, a value much superior than the 1.5% reduction by the planet's disk, and that the hydrogen in its extended atmosphere has velocities ranging from -130 to +100 km s$^{-1}$ . They concluded that the planet is losing neutral hydrogen to space, and subsequent detection of ~13% absorption in OI line and ~7.5% absorption in CII line (Vidal-Madjar et al., 2004) confirmed this conclusion. The fact that heavier elements have also been detected at high altitudes from HD 209458 b, including Si III (Linsky et al., 2010), leads to the conclusion that the atmosphere is in a "blow-off" state (Bourrier and Lecavelier des Etangs, 2013).

After this discovery, the planet HD 189733 b, orbiting a nearby (19.3 pc) K 1.5V type star, was also found to exhibit mass loss (Lecavelier des Etangs et al., 2010; Bouchy et al., 2005); many molecules were discovered in its atmosphere (de Kok et al., 2013; Tinetti et al., 2007; Swain et al., 2008).

Other Hot Jupiters are suspected of losing mass, including WASP-12 b (Li et al., 2010; Maciejewski et al., 2011), and 55 Cnc b (Ehrenreich et al., 2012). In special, the planet 55 Cnc b, with an orbital mean distance of 0.115 AU, slightly violates the formal maximum orbital distance of 0.1 AU for Hot Jupiters. This planet is part of a rich system of five planets (von Braun et al., 2011) orbiting an also nearby (12.3 pc) K0 IV-V type star.

This work explores the possibility of the formation of molecules along the atmospheric mass loss of Hot Jupiters, both oxygen bearing and carbon-chain ones, in a further out region not covered by the literature. We conducted a series of chemical and photochemical reaction simulations, using a network of 56 ions/molecules and 566 reactions retrieved from the UMIST database (McElroy et al., 2013). All the parameter values used for HD 209458b, such as hydrogen density, velocity of gas expansion, temperature and ion/atom abundances, were taken from the literature.

In Section 2 we explore the models of mass loss from Hot Jupiters, in order to acquire parameters on the environment where the molecules are formed and destroyed, including radiation field, C/O ratio, temperature, velocity and number density radial profiles. We elaborate on known chemical and photochemical pathways to the formation and destruction of carbon chain molecules, focusing on benzene and PAHs; then we combine the available physical, chemical and photochemical data, as well as mass loss models, to build the simulations; in Section 3 the results from the simulations are presented and discussed. In Section 4 we draw some conclusions.

## 2. Methods

### 2.1 Mass loss models

Mass loss models of Hot Jupiters are complex, because there are many important and sometimes poorly understood or quantified factors. The incident XUV radiation (X ray and extreme UV), responsible for the heating and swelling of the atmosphere, is not directly measured, since the EUV is strongly absorbed by interstellar atomic and molecular hydrogen; the variable stellar wind also plays a role, as well as tidal forces (Garcia-Muñoz, 2007), interaction of stellar and planetary magnetic fields (Strugarek et al., 2014; Lanza, 2013) and magnetohydrodynamics (Tanaka et al., 2014). Figure 1 illustrates the mass loss from a Hot Jupiter. Until now, no detailed mass loss model in the literature takes into account all these variables.

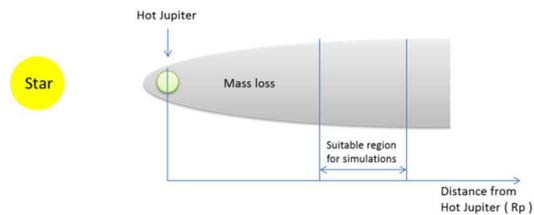

Fig. 1. Illustration (not in scale) of the mass-loss of the atmosphere of a Hot Jupiter. The suitable region for simulations in our study is from 6.9 to 27.5 $R_p$ , where $R_p$ is the radius of the planet.

Most of the mass loss models in the literature for Hot Jupiters like HD 209458b and HD 189733 b predict a hydrogen mass loss rate of $10^9$–$10^{11}$ g s$^{-1}$ (Lammer et al., 2003; Lecavelier des Etangs et al., 2008; Garcia Muñoz, 2007; Penz et al., 2008; Guo 2011, 2013; Murray-Clay et al**.,** 2009**;** Koskinen et al., 2013a, 2013b). Garcia-Muñoz (2007) modelled the physical and chemical aeronomy of HD 209458b, and concluded that beyond a few planetary radii all elements are strongly ionized, assuming that the emission spectrum of the star is identical to the Sun's, and using a network of 223 reactions involving 46 species.

More recently, Bourrier and Lecavelier des Etangs (2013) have developed a 3D particle code in order to model hydrogen atmospheric escape specificallyin the exosphere (altitude higher than ~ 3 Rp) of HD 209458b and HD 189733b, which takes into account radiation pressure, ionization, self-shielding and stellar wind interactions. The fact that their model consider only neutral hydrogen does not exclude it to be used for heavier species (Bourrier et al., 2014, Koskinen et al. (2013b).

As for HD 209458b, Bourrier and Lecavelier des Etangs (2013) conclude that radiation-pressure acceleration account for escape velocities of up to 130 km s$^{-1}$; for HD 189733b, measured velocities up to 230 km s$^{-1}$ require additional acceleration mechanisms, and interaction with stellar wind protons is suggested. In addition, the mass loss from HD 189733b is apparently not steady (Lecavelier des Etangs et al., 2012).

Table 1 summarizes results of recent theoretical models on the mass loss from HD 209458b, namely, velocity, temperature and density radial profiles, which were used in our chemical and photochemical reaction simulations. Except for the model of Bourrier and Lecavelier des Etangs (2013), all the others are hydrodynamic models valid for the thermosphere up to the exobase. In this context it is also important to note that observations by Vidal-Madjar et al. (2013) indicates neutral hydrogen velocities between 40 km s and 130 km s in the exosphere, which is our region of interest, since it is in the exosphere that temperature drops below 2000 K, an upper limit above which many of the molecules studied undergo thermal dissociation

**2.2 Photochemical Rate Constant Corrections**

The reaction constants of the photochemical reactions in the UMIST database are estimated for the interstellar radiation field (McElroy et al., 2013), so they have to be corrected for the environment of Hot Jupiters, where the UV radiation field is many orders of magnitude higher. In order to estimate the correction factor for the mass loss of HD 209458 b and other Hot Jupiters of interest, we modeled the radiation field of the stars, considering them as black bodies, and integrating through the FUV range, extending from 6 to 13.6 eV. The resulting fluxes in photons cm$^{-2}$ s$^{-1}$ are then divided by the ISM UV flux of $10^8$ photons cm$^{-2}$ s$^{-1}$ (Boechat-Roberty et al., 2009), giving the estimated multiplicative correction factor, $C$. By using a black body model for the UV region of a Sun-like star, which is the case of HD 209458, the resulting FUV photon flux is expected to be conservative (higher than the actual), since the actual spectrum is expected to contain several absorption lines (de Pater and Lissauer, 2001). For HD 209458b, the estimated correction factor was $10^9$. This value is conservative, when compared to the integrated and rescaled stellar spectra shown by France et al. (2010) and Vidal-Madjar et al. 2004 (including line emissions), which give correction factors of $5 \times 10^6$ and $10^7$ respectively. Even if we add the estimated continuum contribution from 170 nm to 206 nm, the correction factor from Vidal-Madjar et al. 2004 reaches $5 \times 10^7$. A more realistic correction factor would also require the contribution of X-ray (1 < λ < 100 Å) and EUV (100 < λ < 912 Å) radiation, generated by non-thermal processes in the coronae of stars, which is powered by the stellar rotation, which in turn decreases with age. But since EUV radiation is mostly absorbed by interstellar atomic and molecular hydrogen, correlations based on observed X-ray luminosities by X-ray telescopes like XMM-Newton, and EUV luminosities based on scarce EUV luminosities available from the Sun and a few other stars, have been built (Sanz-Forcada et al., 2011).

**Table 1**
Results from theoretical models of mass loss from HD 209458 b; * private communication; $R_p$ is the planet radius

| Reference | HI Velocity (km s$^{-1}$) profile | Temperature (K) profile | HI Density ( cm$^{-3}$ ) profile |
|---|---|---|---|
| Bourrier and Lecavelier des Etangs (2013) | 0 to 100 in 5 h | | $1.3 \times 10^6$ at 2.8 $R_P$ to $1 \times 10^1$ at 20 $R_P$ * |
| Guo (2013,2011) | 0 to 18 between 1 and 5 $R_p$ | 6300 at 3.5 $R_p$ to 5000 at 5 $R_p$ | $1.6 \times 10^6$ at 3Rp to $2 \times 10^5$ at 5Rp |
| Koskinen et al. (2013a) | 10 to 30 between 5 and 20 $R_p$ | 8000 at 4 $R_p$ to 3000 at 11 $R_p$ | $10^7$ at 3 $R_p$ to $10^6$ at 5 $R_p$ |
| Murray-Clay et al. (2009) | 0 to 15 between 1 and 5 $R_p$ | 5000 at 3.5 $R_p$ to 3400 at 5 $R_p$ | $1.3 \times 10^6$ at 3 $R_p$ to $1.6 \times 10^5$ at 5 $R_p$ |
| Penz et al. (2008) | 3 to 8 between 3 and 5 $R_p$ | 4100 at 3 $R_p$ to 2500 at 5 $R_p$ | $10^7$ at 3 $R_p$ to $7 \times 10^5$ at 5 $R_p$ |
| Garcia-Muñoz (2007) | 10 to 20 between 5 and 15 $R_p$ | 6000 at 5 $R_p$ to 3000 at 15 $R_p$ | |

**Table 2**
Integrated luminosities in wavelength ranges: $L_X$, at X-ray (5-100 Å) and $L_{EUV}$, at EUV (100-920 Å) of the host stars and the respective fluxes $F_X$ and $F_{EUV}$ incident on selected Hot Jupiter (Sanz-Forcada et al. 2011)

| Planet | log $L_X$ (erg s$^{-1}$) | log $L_{EUV}$ (erg s$^{-1}$) | log $F_X$ (erg s$^{-1}$ cm$^{-2}$) | log $F_{EUV}$ (erg s$^{-1}$ cm$^{-2}$) | $F_X$ (photons cm$^{-2}$ s$^{-1}$) | $F_{EUV}$ (photons cm$^{-2}$ s$^{-1}$) |
|---|---|---|---|---|---|---|
| 55 Cnc b | 26.65 | 27.49 | 1.08 | 1.92 | ~ $10^{10}$ | ~ $10^{12}$ |
| HD209458 b | < 26.40 | < 27.74 | < 1.60 | < 3.29 | ~ $10^{11}$ | ~ $10^{14}$ |
| HD189733 b | 28.18 | 28.48 | 3.73 | 4.03 | ~ $10^{13}$ | ~ $10^{14}$ |

Table 2 shows measurements of X-ray (5-100 Å) and EUV (100-920 Å) luminosities by XMM-Newton and Chandra observatories for three Hot Jupiter host stars, which were used by Sanz-Forcada *et al.* (2011). As expected, the old star 55 Cnc (~ 10.2 Gyr) presents a weaker XUV (X-ray + EUV) luminosity than that of HD209458 (~ 4 Gyr), and much weaker than that of HD189733 (~ 0.6 Gyr), These data will be used, for the simulation scenarios in section 4.

**2.3 C/O ratios for Hot Jupiter**

The study of C/O ratio in Hot Jupiters atmospheres is important for theories of planet formation and migration (Johnson et al., 2012; Öberg et al., 2011; Stevenson and Lunine, 1988; Ida and Lin, 2004; Pollack et al., 1996). The condensation temperature of the ensemble of available volatile and refractory species (whose relative abundances is defined by the C/O ratio), will dictate the radial density profile of ices and dust, which form the cores of gas giant planets and the bulk of rocky ones respectively. It is also fundamental for the modelling of Hot Jupiter atmospheres, since the C/O in hydrogen-rich atmospheres will define the relative abundances of several molecules and the temperature structure.

Some studies of transiting Hot Jupiters have indicated considerable deviations from solar abundances (Line et al., 2014; Madhusudhan, 2012; Madhusudhan et al., 2011), with C/O ratios significantly higher than the solar value. Madhusudhan (2012) has proposed C/O ratio and irradiation (or temperature) as the two main factors dictating atmospheric chemistry and structure.

Another approach to C/O ratio in exoplanets in general is the study of the C/O ratio of their host stars, which, along with metallicity and refractory element abundance estimates, provide a more robust starting point for inferences about formation probability, composition and, perhaps, the optimal locus of their planets (Nissen, 2013; Ghezzi et al, 2010; Meléndez et al., 2009; Fischer and Valenti, 2005; Pinotti et al., 2005). But this approach is subject to debate; for example, Delgado Mena et al. (2010) have estimated a high C/O ratio for 55 Cnc (1.12), which led to the supposition that the interior of the super-earth 55 Cnc e may also be carbon-rich (Madhusudhan et al., 2012). But the very high C/O ratio of 55 Cnc have been disputed by Teske et al. (2013), who found a value of 0.78, on the basis that the O abundance evaluation requires a much more detailed analysis than is usually made.

More recently, Teske et al. (2014) published a list of C/O ratios of stars with transiting Hot Jupiters, which indicate that previously-measured exoplanet host star C/O ratios may have been overestimated. Their work also supports evidence of increase in C/O with Fe/H (Nissen, 2013).

Line et al. (2014) have made an analysis of C/O ratio for a number of transiting exoplanets based on secondary eclipse spectroscopy. They have found a value of C/O = 1 for HD 209458b, which is in line with the result of C/O ≥ 1 by Madhusudhan and Seager (2009), although the C/O ratio of the parent star HD 209458 is apparently low (Johnson et al., 2012).

The objective of this development is not to exhaust the ongoing discussion on C/O ratio of stars and their planets, but to justify our assumptions on C/O ratios used in the simulations, based on observational evidence. From the previous exposition we conclude that it is reasonable to suppose that some Hot Jupiters, beyond HD209458b, may indeed have a high C/O ratio, approaching or even surpassing C/O = 1, by carbon enrichment during planetary formation. Therefore, we will use a range of C/O ratios for HD209458b and hypothetical Hot Jupiters that exhibit similar mass loss profiles.

As a result of the evidence shown above, we decided to use C/O=1.0 in our base simulations, as well as to explore how the results are affected by changing the C/O ratio.

Finally, it is worth mentioning that Jupiter itself may have a higher C/O ratio than that of the Sun (Mousis et al., 2012; Wang and Lunine, 2012). The main uncertainty factor regarding the evaluation of C/O ratios of solar system's giant planets is $H_2O$ condensation down to deeper layers due to low atmospheric temperatures (Wang D. et al., 2015).

**2.4 Building the simulations**

In order to simulate scenarios of molecular formation along the mass loss of Hot Jupiter HD 209458 b, we used a 1D *general* chemical reaction model, based on the molecule balance on the element volume $Adr$, where $A$ is the area perpendicular to the radial velocity $v$. The model is represented by

$$\frac{d}{dr}[\rho v(x_i)] = \sum \Omega_n \qquad (1)$$

where $\rho$ is the hydrogen number density (cm$^{-3}$), $v$ is the radial velocity of the mass loss (cm s$^{-1}$), $\Omega_n$ is the reaction rate (cm$^{-3}$ s$^{-1}$) of any reaction $n$ (from the 566 reactions used) which produces or consumes the species $i$, and $x_i$ is the abundance of the species $i$ *relative to hydrogen*. The parameters $v$, $x_i$ and $\rho$ are function of $r$ ($x_i=x_i(r)$, $\rho= \rho(r)$ and $v= v(r)$). Notice that the equation does not include a "*squared-r*" factor, since the mass loss from Hot Jupiters is *not* an expanding spherical shell of gas. The radial profiles of ρ and v and will be taken from the literature (see Section 2) and used by the differential equation solver.

The rate constants for the reactions $\Omega_n$ are defined according the UMIST database formalism (McElroy et al., 2013):

$$k_n = \alpha_n \left(\frac{T}{300}\right)^{\beta_n} exp\left(\frac{-\gamma_n}{T}\right) \text{ cm}^3 \text{ s}^{-1} \quad (2)$$

for two-body reactions, where T (K) is the gas temperature ($T = T(r)$) and $\alpha_n$, $\beta_n$ $\gamma_n$ and are parameters available in the database; for photoreactions, the rate constant is given by:

$$k_{phot,n} = \alpha_n exp(-\gamma_n A_V) \text{ s}^{-1} \quad (3)$$

Where $\alpha_n$ represents the rate coefficient in the unshielded interstellar ultraviolet radiation field, and $\gamma_n$ and $A_V$ are parameters accounting for dust extinction in the UV and visible respectively.

We have also included direct cosmic-ray ionization reactions and cosmic-ray-induced photoreactions, although their very low rate constants ($10^{-15} - 10^{-17}$ s$^{-1}$ – UMIST database) do not have a significant impact on the main results, since the time scale of the simulations is measured in hours ($10^3$ s) and estimates of molecular dissociation along interplanetary distances is measured in days ($10^4$ - $10^5$ s). However, direct cosmic-ray ionization reactions and cosmic-ray-induced photoreactions could play a role in the abundance of trace species.

Since the temperature also depends on the radial profile of the mass loss ($T=T(r)$), it follows that the rate constants will also vary radially in the simulations ($k_n=k_n(r)$ and $k_{phot,n}=k_{phot,n}(r)$).

For most of the *n* photochemical reactions used in the simulations, the rate constants ($k'_{phot,n}$, s$^{-1}$) will be corrected according to the factors estimated in section 2.1, that is,

$$k_{phot,n} = k'_{phot,n} * C(r)$$

where $C(r)$ is the correction factor; since the radiation field of the host star decreases with the squared distance ($C \propto 1/r^2$), and the reaction locus is of considerable radial length (~ $10^6$ km) in a region very near the star, *C* is relatively sensible to distance. Therefore, we have built in the simulation correction factors that depend on the distance from the star.

We have run the simulations by using 566 reactions from the UMIST database, involving 56 species containing H, He, C and O. Table 4 shows the list of the species considered. The simulations were modeled using MATLAB, coupled with the solver DASSLC (Differential Algebraic System Solver in C) adapted for use in MATLAB (Secchi, 2010), which integrates the set of 53 differential equations (HI, HeI and $H_2$ profiles are predetermined).

The simulations require density, temperature and velocity radial profiles for hydrogen along the mass loss. For the case of HD 209458b, we assumed the density profile of Bourrier and Lecavelier des Etangs (2013), since it has been developed specifically for the exosphere; the velocity profile has been taken from Vidal-Madjar et al. (2003). Table 4 summarizes our profile definitions; the initial radius chosen for the simulations was 6.9 $R_p$, where the temperature is around 2000 K (estimated by extrapolating the hydrodynamic models), the limit where water molecule suffers thermal dissociation; the end radius was 27.5 $R_p$, where the hydrogen density is low enough so that chemical reaction rates start to be surpassed by photodissociation reaction rates. The resulting radial length spans ~ 2 x $10^6$ km. In order to run the simulations, a set of initial abundances of species (at 6.9 $R_p$) is required; we have taken these from Koskinen et al. (2013a), who used solar abundances. Abundance data from Koskinen et al. (2013) reach 5 $R_p$, so we have underestimated the initial ion abundances relative to atomic abundances. However, in section 5 we will show that the main results are not very sensitive to ionization degree. Table 5 shows the values of initial abundances at 6.9 $R_p$; for C/O different from the solar value, the abundances of the C and O species were corrected proportionally, maintaining the total atom number (hence the metalicity).

The values of electron density in the simulations are centered around 5 x $10^{-3}$ times the hydrogen density, or ~ $10^3$ cm$^{-3}$, which is consistent with calculations of electron densities (Koskinen et al., 2014) at the higher atmosphere of HD 209458 b ($10^{12}$-$10^{14}$ m$^{-3}$ at $10^{-10}$ bar), once allowing for the dilution along the mass loss.

The abundances of chemical species along the atmospheric layers of HD 209458 b have been modeled (Moses *et al*., 2011; Garcia-Muñoz 2007; Koskinen, 2013a) and the results generally agree that the $H_2$ molecule abundance drops steeply at $10^{-6}$ – $10^{-7}$ bar, where hydrogen atoms become the dominant species. Moses et al. (2011) includes both disequilibrium thermochemistry and photochemistry in their model in order for it to become more realistic, and conclude that the radical OH efficiently attacks

H$_2$. They assume that molecular and eddy diffusion are the relevant vertical transport mechanism, and that the atmospheric profile exhibits thermal inversion, which would pose an obstacle to possible convective cells toward the thermosphere.

Table 3
Species considered in the simulations

| Species with H | Species with He | Species with C | Species with O | Species with H and C | Species with H and O | Species with C and O | Species with H, C and O |
|---|---|---|---|---|---|---|---|
| H$^+$ | He$^+$ | C$^+$ | O$^+$ | CH$^+$, CH, CH$^-$, CH$_2$$^+$, CH$_2$, CH$_3$$^+$, CH$_3$, CH$_4$$^+$, CH$_4$, C$_2$H$^+$, C$_2$H, C$_2$H$^-$, C$_2$H$_2$$^+$, C$_2$H$_2$, C$_2$H$_3$$^+$, C$_2$H$_3$, C$_2$H$_4$$^+$, C$_2$H$_4$, C$_3$H$_2$, C$_3$H$_3$$^+$, C$_3$H$_3$, C$_3$H$_4$$^+$, C$_4$H$_2$$^+$, C$_4$H$_3$$^+$, C$_5$H$_3$$^+$, C$_6$H$_5$$^+$ C$_6$H$_6$, C$_6$H$_6$$^+$, C$_6$H$_7$$^+$ | OH$^+$ OH OH$^-$ H$_2$O$^+$ H$_2$O H$_3$O$^+$ | CO$^+$ CO CO$_2$$^+$ CO$_2$ | HCO HCO$^+$ H$_2$CO |
| H$^-$ | He | C$^-$ | O$^-$ | | | | |
| H | | C | O | | | | |
| H$_2$$^+$ | | | | | | | |
| H$_2$ | | | | | | | |
| H$_3$$^+$ | | | | | | | |

Table 4
Profiles of velocity, density and temperature used in the simulations for HD 209458 b. 1 R$_p$ = 1.38 R$_J$

| Variable | Value at 6.9 Rp | Value at 27.5 Rp | Obs |
|---|---|---|---|
| Temperature (K) | 2000 | 500 | Power law with α= -1 |
| Density (cm$^{-3}$) | 5.0 x 10$^3$ | 1.0 x 10$^1$ | Power law with α= -2.8 |
| Velocity (km s$^{-1}$) | 40 | 130 | Linear |

Table 5
Abundances relative to neutral hydrogen (HI) used as the starting point of the simulations; solar metalicity

| C/O | 0.68 | 0.54 (solar) | 0.75 | 1.00 | 1.25 | 1.50 | 2.00 |
|---|---|---|---|---|---|---|---|
| H$^+$ | 1.5 x 10$^0$ | 1.5 x 10$^0$ | 1.5 x 10$^0$ | 1.5 x 10$^0$ | 1.5 x 10$^0$ | 1.5 x 10$^0$ | 1.5 x 10$^0$ |
| C | 2 x 10$^{-4}$ | 1.75 x 10$^{-4}$ | 2.14 x 10$^{-4}$ | 2.5 x 10$^{-4}$ | 2.78 x 10$^{-4}$ | 3 x 10$^{-4}$ | 3.33 x 10$^{-4}$ |
| C$^+$ | 8 x 10$^{-4}$ | 7.00 x 10$^{-4}$ | 8.55 x 10$^{-4}$ | 1.0 x 10$^{-3}$ | 1.11 x 10$^{-3}$ | 1.2 x 10$^{-3}$ | 1.33 x 10$^{-3}$ |
| O | 5 x 10$^{-4}$ | 5.40 x 10$^{-4}$ | 4.79 x 10$^{-4}$ | 4.17 x 10$^{-4}$ | 3.70 x 10$^{-4}$ | 3.33 x 10$^{-4}$ | 2.78 x 10$^{-4}$ |
| O$^+$ | 1 x 10$^{-3}$ | 1.08 x 10$^{-3}$ | 9.58 x 10$^{-4}$ | 8.34 x 10$^{-4}$ | 7.40 x 10$^{-4}$ | 6.66 x 10$^{-4}$ | 5.56 x 10$^{-4}$ |
| He | 5 x 10$^{-2}$ | 5 x 10$^{-2}$ | 5 x 10$^{-2}$ | 5 x 10$^{-2}$ | 5 x 10$^{-2}$ | 5 x 10$^{-2}$ | 5 x 10$^{-2}$ |
| He$^+$ | 2 x 10$^{-1}$ | 2 x 10$^{-1}$ | 2 x 10$^{-1}$ | 2 x 10$^{-1}$ | 2 x 10$^{-1}$ | 2 x 10$^{-1}$ | 2 x 10$^{-1}$ |

Table 6
Estimated values of k for photochemical reactions involving C$_6$H$_6$ and C$_6$H$_6$$^+$

| Reaction | σ (cm$^2$) | Estimated k (s$^{-1}$) | | |
|---|---|---|---|---|
| | | HD 209458 b | HD 189733 b | 55 Cnc b |
| C$_6$H$_6$ + FUV → C$_6$H$_6$$^+$ + e$^-$ | 1.5 x 10$^{-17}$ | 3.0 x 10$^0$ | 3.2 x 10$^{-1}$ | 6.3 x 10$^{-2}$ |
| C$_6$H$_6$ + FUV → C$_2$H$_2$ +C$_4$H$_4$ | 1.5 x 10$^{-18}$ | 3.0 x 10$^{-1}$ | 3.2 x 10$^{-2}$ | 6.3 x 10$^{-3}$ |
| C$_6$H$_6$ + X-ray → C$_6$H$_6$$^+$+ e$^-$ | 1.2 x 10$^{-18}$ | 1.2 x 10$^{-7}$ | 1.2 x 10$^{-5}$ | 1.2 x 10$^{-8}$ |
| C$_6$H$_6$ + X-ray → C$_2$H$_2$ +C$_4$H$_4$ | 3.8 x 10$^{-17}$ | 3.8 x 10$^{-6}$ | 3.8 x 10$^{-4}$ | 3.8 x 10$^{-7}$ |

However, they acknowledge that hydrodynamic flow could affect the stratospheric composition, if an atmospheric bulk wind dominates down to the base of the thermosphere. There is observational evidence of strong winds in HD 209458 b down to at least 10$^{-5}$ bar (Snellen et al., 2010), as well as observational evidence of the lack of thermal inversion in the atmosphere of HD 209458 b (Schwarz et al., 2015; Diamond-Lowe et al., 2014). Moreover, France et al. (2010) detected a significant feature in the FUV spectra of HD 209458 b, which may be due to $H_2$ excitation. The feature, at 1582 Å, was observed at +15 (± 20) km s$^{-1}$ in the rest frame of the planet. They admit that the excitation mechanism is unclear, and point out the possibility of excitation by EUV radiation from the star, stressing that the possibility is remote due to attenuation by H atoms in the upper atmosphere. But an alternative interpretation is that the feature at 1582 Å is indeed due to $H_2$ in the thermosphere/exosphere with the velocity of + 15 km s$^{-1}$, which is consistent with values of mass loss velocities discussed in this paper, and where the H density is sufficiently low to allow $H_2$ excitation by EUV radiation from the star. Finally, the detection of Mg I, a heavy atom, escaping from the planet at several radii (Vidal-Madjar et al., 2013) gives more credence to the case for atmospheric hydrodynamic escape. Furthermore, there is observational evidence of the existence of an optically thin dust layer (carbonaceous, TiO, VO, silicates) in the outer atmosphere of HD 209458b (Richardson et al. 2007, Burrows et al. 2008), which could shield the $H_2$ molecules from the stellar radiation field, thus extending their survival through the thermosphere and exosphere.

Given these evidences and the extreme and poorly understood conditions of the atmosphere of Osiris, we assume in this work that 3D hydrodynamic flow allows residual $H_2$ molecules to reach the thermosphere and exosphere, taking part of the chemical/photochemical reaction simulations with abundances ranging from $10^{-1}$ to $10^{-6}$ relative to HI.

The values of $A_v$ used in the simulations will be set between 0.5 and 1.5, consistent with the total hydrogen column density from 6.9 $R_p$ and 2 $R_p$ in the mass loss references used in this work, and with the formula $A_v = N_H / 1.6x10^{21}$ (Wakelam et al., 2013). Although the formula is fit in principle for the ISM, the metallicity of the environment studied here is far greater than the metallicity of the ISM, so the extinction in the UV from atoms would be more intense; if we also consider the possible presence of dust described above, we conclude that the estimate of $A_v$ is probably conservative, and, although a precise estimation is not possible, the reader will be able to appreciate the effect of this parameter on the results.

For photoreactions involving benzene and the benzene cation $C_6H_6^+$, we developed a more detailed treatment (see 3.1.1), since we have information on cross sections for incident X-ray and EUV (Section 2) which are practically absent in the ISM, and also the X-ray and EUV fluxes. Table 7 summarizes the results for $k$ for photochemical reactions involving $C_6H_6$ and $C_6H_6^+$, for three Hot Jupiters.

These reaction rates are also modulated by a factor which decreases with the squared distance. A rapid inspection on Table 6 shows an important conclusion, that is, although XUV photons (X-ray + EUV) are more energetic than FUV photons, the contribution of the former in the ionization and dissociation processes is irrelevant compared to the latter, in the context of Hot Jupiters, and that is due to the high difference in photon flux. For HD 209458 b, for example, while the estimated FUV photon flux is 2.0 x $10^{17}$ cm$^{-2}$ s$^{-1}$, the EUV and X-ray fluxes are $10^{14}$ cm$^{-2}$ s$^{-1}$ and $10^{11}$ cm$^{-2}$ s$^{-1}$ respectively.

## 3 Results and Discussion

### 3.1 Carbon chain molecules

Depending on the C/O ratio used in simulation, carbon-chain molecules might be preferentially formed, so we included in the simulations carbon reactions which lead to the formation of Benzene, the precursor of Polycyclic Aromatic Hydrocarbon (PAH).

The PAH hypothesis was formulated, 30 years ago (Léger and Puget, 1984; Allamandola et al., 1985; Tielens, 2011), in order to account for strong mid-IR emission features in the spectra of a great variety of astronomical objects. Theoretical, experimental and observational studies have not only corroborated it, but also developed techniques to understand physical and chemical characteristics of the sources, based on the complexity of the PAH emission bands. This complexity is caused by the fact that the family of PAHs is extensive, both due to the carbon number and in molecular configurations. Complexity variables include mergers of aliphatic and aromatic groups (Kwok and Zhang, 2013), multiple ionization states (Sloan et al., 1999; Maaskant et al., 2014), the presence of hetero-atoms such as nitrogen in the aromatic ring (Hudgins et al., 2005), and clusters/complexes with iron (Simon et al., 2011).

The main source of interstellar PAHs is thought to be evolved circumstellar environments (Cherchneff, 2011), which have the necessary formation conditions, that is, an initial carbon, hydrogen rich chemical composition, high densities and high temperatures. Although UIR emission bands have been detected in some of these objects, and attributed to PAH excitation through absorption UV radiation emitted by a companion star, the lack of detection of UIR emission bands from most of the carbon stars is probably due to the lack of exciting radiation. For example, carbon rich AGB stars have surface temperatures of the order of 3,000 K, too low for the emission of a substantial UV radiation field.

However, the mass loss from Hot Jupiters with high C/O ratios might be a new possibility of PAH forming region where the UV radiation field is strong enough to excite the molecules, allowing the observation of IR emission features, and, perhaps, the identification of specific PAH molecules.

### 3.1.1 Formation and destruction of $C_6H_6$

In the Solar System, only the simplest aromatic molecule, benzene, was identified in the atmosphere of the planets Jupiter and Saturn (Kim et al., 1985; Bézard et al., 2001) and in Titan's atmosphere (Coustenis et al., 2003; Vinatier et al., 2007). Outside the Solar System, benzene was again the only unambiguously identified PAH, a discovery made by Cernicharo et al. (2001) in the proto-planetary nebula CRL 618.

Benzene formation is the rate-limiting step in chemical pathways to PAHs. Therefore, laboratory studies of benzene formation, as well as simulations, have been extensively conducted in order to estimate benzene abundances in astronomical objects. Woods et al. (2002) simulated the formation of benzene in CRL 618, using a route of ion-molecule reactions which is based on high abundances of $HCO^+$ and $C_2H_2$ and a high flux of ionizing photons from the hot central star. The final step of this path is the dissociative recombination of the $c$-$C_6H_7^+$ ion, which is also thought to be the final step leading to the formation of benzene in Titan's atmosphere (Vuitton et al., 2008). Experimental studies by Hamberg *et al.* (2011) conclude that dissociative electron recombination of $c$-$C_6H_7^+$ can be regarded as a feasible final step in the synthesis of benzene in the interstellar medium and planetary atmospheres.

The chemical pathways used in simulations of benzene formation in the circumstellar environments of carbon-rich AGB stars are based on studies of acetylenic flames (Cherchneff, 2011), and involve radicals, with the assumption of negligible ionization by radiation (see Table 7). However, as shown in Section 2.1, the environment of mass loss from Hot Jupiters is dominated by a strong UV radiation field, and therefore the ion-molecule mechanism proposed by Woods et al. (2002) was chosen for our simulations (see Table 8).

**Table 7 --**. Benzene formation mechanisms in acetylenic flames. Source: Cherchneff (2011)

| |
|---|
| $C_3H_3 + C_3H_3 \rightarrow C_6H_6$ |
| $C_3H_3 + C_3H_3 \rightarrow C_6H_5 + H$ |
| $1$-$C_4H_3 + C_2H_2 \rightarrow C_6H_5$ |
| $1$-$C_4H_5 + C_2H_2 \rightarrow C_6H_6 + H$ |

The understanding of photoionization and dissociative photoionization of the benzene molecule is important for the estimation of its net production rate (formation – destruction) in astrophysical environments; Boechat-Roberty et al. (2009) have studied the UV and soft X-ray photons interaction with benzene molecule, using synchrotron radiation coupled with time of flight mass spectrometry. They confirmed the relative high stability of $C_6H_6$ under UV radiation, where the ion $C_6H_6^+$ is preponderant over few different fragments from the destruction of $C_6H_6$, and its vulnerability under soft X-ray radiation, which produces many and abundant fragments.

**Table 8**. Formation pathways of benzene in CRL 618, according Woods et al. (2002).

| |
|---|
| $HCO^+ + C_2H_2 \rightarrow C_2H_3^+ + CO$ (1) |
| $C_2H_3^+ + C_2H_2 \rightarrow C_4H_3^+ + H_2$ (2) |
| $C_4H_3^+ + C_2H_2 \rightarrow c$-$C_6H_5^+ + h\nu$ (3) |
| $c$-$C_6H_5^+ + H_2 \rightarrow c$-$C_6H_7^+ + h\nu$ (4) |
| $c$-$C_6H_7^+ + e^- \rightarrow c$-$C_6H_6 + H$ (5) |

This experimental work has allowed the determination of the photoionization ($\sigma_{ph\text{-}i}$ $(E)$) and photodissociation ($\sigma_{pf\text{-}d}$ $(E)$) cross sections as a function of the photon energy $E$, of benzene, in the UV and X-ray ranges. Integrating the respective values of cross sections $\sigma$ $(E)$ (cm$^2$) and the photon flux f$(E)$ (photons cm$^{-2}$ s$^{-1}$) in a range of photon energies, from $E_1$ to $E_2$, it is possible to estimate the photodissociation and photoionization rate constants:

$$k_{ph-i} = \int_{E_1}^{E_2} \sigma_{ph-i}(E) f(E) dE$$

$$k_{ph-d} = \int_{E_1}^{E_2} \sigma_{ph-d}(E) f(E) dE$$

Benzene abundance results suggest the lack of carbon-chain molecules formation along the mass loss of HD 209458b. For $A_v$ =1.0 and C/O = 1.0 the peak mixing ratio (relative to HI) is 10$^{-54}$. Variations of Av and C/O do not improve the scenario. The main reason for the absence of benzene in significant abundances is that acetylene, which participates in three out of five steps in the reaction chain leading to benzene; it does not reach significant mixing ratio either (< 10$^{-36}$). It should also be noted that benzene formation is the result of many reactions, which translates to a low overall reaction rate. These results agree with previous studies that indicate the lack of formation of carbon-chain species in these environments.

**3.3 Molecules containing oxygen and hydrogen**

The fractional abundance profiles (relative to HI) for simple species containing oxygen and hydrogen are distinctively improved over the ones for hydrocarbons, as Figures 2 show (distance is relative to the starting point of simulations, at 6.9 $R_p$). The peak fractional abundance for $H_2O^+$ reaches $10^{-10}$ and the ion $OH^+$ reaches a remarkable peak value of $10^{-7}$, with an estimated total column density of $3 \times 10^7$ cm$^{-2}$. Small species containing carbon and oxygen (Figure 3) exhibit fractional abundance peak values between $10^{-14}$ and $10^{-19}$. In general, the simulation results are insensitive to the initial ionization ratios CII/CI and OII/OI, so our assumption of ratios in Table 5 is robust.

Figure 4 shows the results for $OH^+$ when all carbon and oxygen are initially in the form of CII and OII, for different $H_2$ abundance profiles; for a $H_2$ profile of $10^{-1}$ to $10^{-2}$, which is used for all the results shown previously, the peak fractional abundance of $OH^+$ remains in the level of $10^{-7}$. However, the peak fractional abundance is very sensitive to $H_2$ fractional abundance profile: the $OH^+$ peak abundance drops to $10^{-10}$ for a $H_2$ fractional abundance radial profile ranging from $10^{-5}$ to $10^{-6}$, causing the estimated total column density of OH to drop to $3.3 \times 10^3$ cm$^{-2}$.

Therefore, a spectroscopic detection of $OH^+$ during the transit of HD209458 b would not only be a discovery of a new species in this environment, but also a strong indication of the presence of $H_2$ in significant abundances along the mass loss.

Figure 5 shows how the results for $OH^+$ are altered by changing the value of $A_v$, which is, as discussed before, a parameter subject to uncertainty. For the most favorable case of $A_v = 1.5$, the peak fractional abundance of $OH^+$ reaches $10^{-6}$, and its estimated total column density, $6 \times 10^7$ cm$^{-2}$; for the least favorable case ($A_v = 0.5$) the results are $10^{-7}$ and $1 \times 10^7$ cm$^{-2}$ respectively. So, the total column density does not change more than one order of magnitude with 50% variation of $A_v$.

Figure 6 shows the effect of C/O ration on the $OH^+$ profile. This parameter has less impact than the one of A, and the peak fractional abundance stays above $4 \times 10$ for any value of C/O ranging from solar and 1.5., with total column densities of $4.3 \times 10^7$ and $2.7 \times 10^7$ cm$^{-2}$ respectively.

Finally, Figure 7 shows the simulation results for variations in the correction factor C. As discussed before, our standard value of $10^9$ is probably conservative, and values down to $\sim 10^8$ are a possibility for HD 209458b. The chosen values of $C$ range from $10^8$ to $5 \times 10^9$, with estimated total column densities of $OH^+$ ranging from $6.5 \times 10^7$ cm$^{-2}$ and $1.1 \times 10^7$ cm$^{-2}$ respectively. Therefore, if we take into account the uncertainty of C, the total column density of $OH^+$ still remains in the $10^7$ cm$^{-2}$ level. An alternative interpretation of Figure 7 is that of moving the planet closer or away from the star; the correction factor will be proportional to the radiation flux, which scales to $1/r^2$, where $r$ is the orbital distance. This means, for example, that the simulation results for $C = 2.5 \times 10^8$ would be that for a planet similar to Osiris, orbiting an identical star, but with an orbital distance twice that of the original value. The conclusion would be that the total column density of $OH^+$ would not be sensibly altered by considerable orbital distance variation. However, this extrapolation has to be taken with extreme reserve, since the alteration of the radiation flux would alter the mass loss profiles of density, velocity and temperature, which remains fixed in the simulations. But, considering also that the new profiles for more distant planet would be more favorable for molecular formation (lower velocities and temperatures, higher densities), we could speculate that the column densities for our simulations would be minimum values for more distant planets similar to Osiris and orbiting identical stars. Another speculation exercise allowed by the simulations is that lower/higher C values correspond to a planet orbiting later/earlier type stars at the same orbital distance. This scenario has to be taken with even greater reserve, since there are, other factors that add to the alteration of mass loss profiles, such as increased stellar activity.

The abundance of all molecules start to drop at the distance of $\sim 2 \times 10^{10}$ cm; at this point, the plasma density is low enough so that the dissociation photoreactions start to take over. In order to estimate the abundances we use the results of our chemical reaction model of HD 209458 b, and equation (4) for numerical integration, considering that the species are only subjected to dissociation from the stellar radiation field:

$$X = X_0 e^{-t k_0 r_0^2 S} \quad (4)$$

where $X_0$ and $X$ are the initial and final abundances, respectively, $k_0$ is the dissociation rate at $r_0$ (0.005 AU + 27.5 $R_p$ = $10^7$ km), $t$ is the time lapse between steps ($t = \Delta r / \bar{v}$), $\bar{v}$ is the mean velocity of the species, and S is the sum of the steps, with each step with a correction factor for the decreasing $k_0$ value with distance:

$$S \sim \sum_{i=0}^{n} \frac{1}{(r_0 + i\Delta r)^2}$$

$$S = \lim_{n \to \infty} \sum_{i=0}^{n} \frac{1}{(r_0 + i\Delta r)^2}$$

The results of the estimates, for values of $\bar{v}$ ranging from 30 to 100 km s$^{-1}$, show that the species are restricted to a radial region of less than $10^7$ km.

Our 1D model is a crude representation of the actual environment where we studied the formation of molecules. Beyond the intrinsic limitations of a 1D model, it uses data which is the result of other simulations, like our choice of parameters (velocity, temperature, number density) from the literature, and the correction factor $C$ which is applied equally to all photochemical reactions. Besides, the fact that the mass loss is almost certainly not constant in time, as observed in HD 189733 b, will certainly have a weight on the results.

But our objective is to probe the possibility of molecular formation, and the order of magnitude of column densities indicate that this possibility exists. Moreover, there are factors not accounted for in this study that could amplify the total column density of OH$^+$: the magnetosphere of the planet (Strugarek et al., 2014; Lanza, 2013), which acts as an inhibitor to dilution of the mass loss, theoretically allowing a less steep radial decline in the number density of ions and consequently sustaining molecular formation in an extended region; and the possible presence of dust particles in the mass loss, synthesized in the lower atmosphere of the Hot Jupiter. These dust particles would have an attenuation effect on the incident XUV radiation, allowing a higher abundance of OH$^+$ and other species along the formation region.

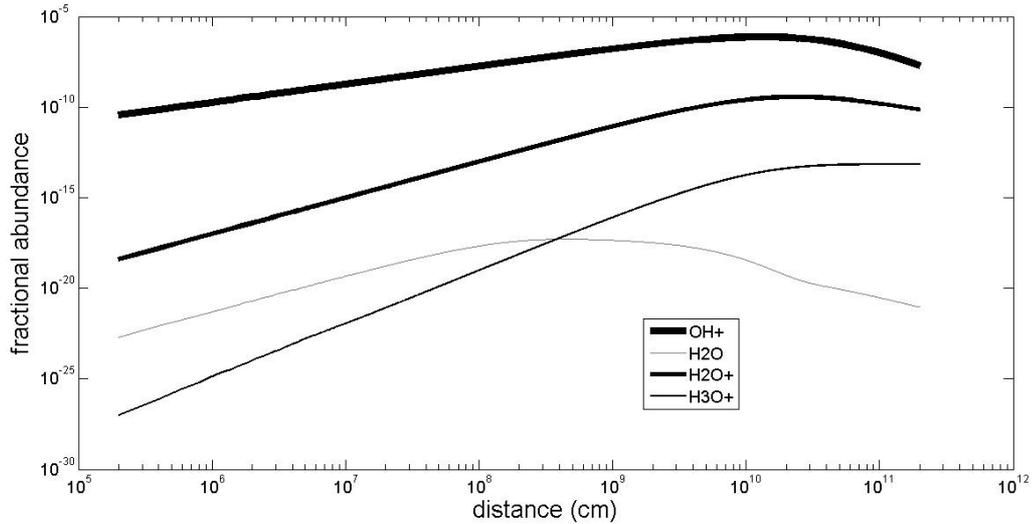

Fig. 2. Fractional abundance of selected species along the atmospheric mass loss of HD 209458 b,; $C=10^9$, $A_v = 1.0$, and C/O = 1.0

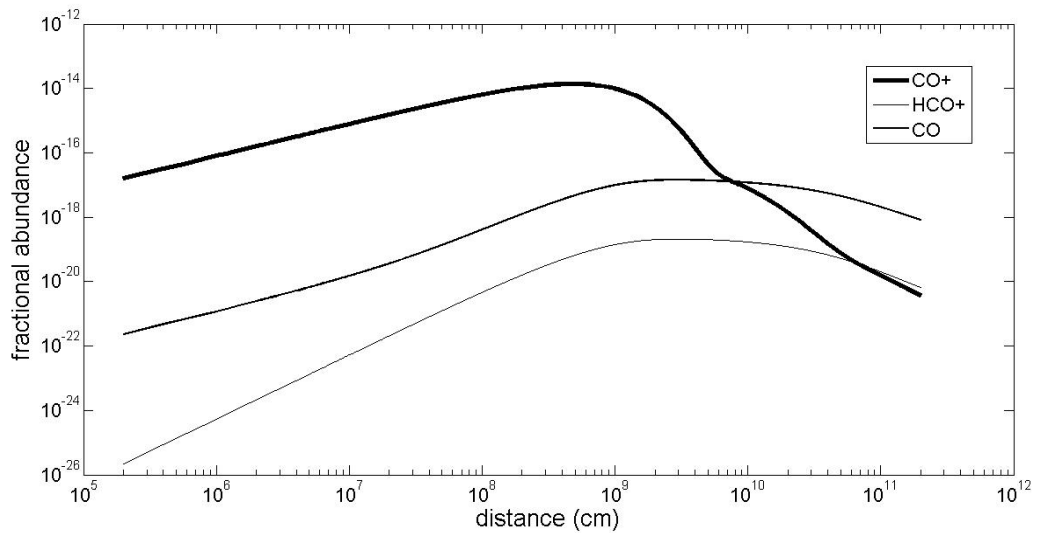

Fig. 3. Fractional abundance for selected species along the mass loss of HD 209458 b,; $C= 10^9$, $A_v = 1.0$, and C/O = 1.0

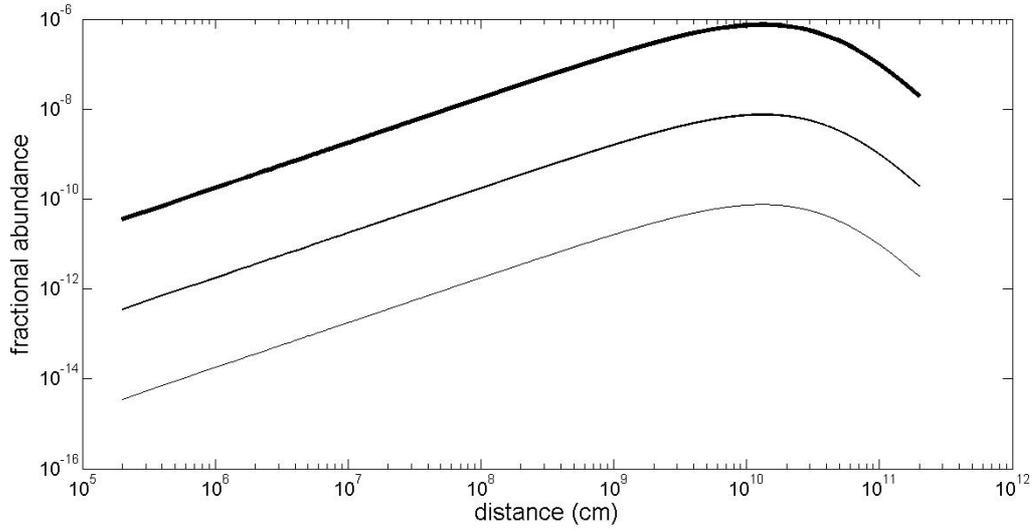

Fig. 4. Results for OH$^+$ fractional abundance along the mass loss of HD 209458 b for three different H$_2$ abundance profiles; thick line: $10^{-1}$ to $10^{-2}$ times the HI abundance; medium line: $10^{-3}$ to $10^{-4}$ times the HI abundance; thin line: $10^{-5}$ to $10^{-6}$ times the HI abundance; Av = 1.0, C/O = 1.0; all carbon ionized to CII, all oxygen ionized to OII at the start of simulations

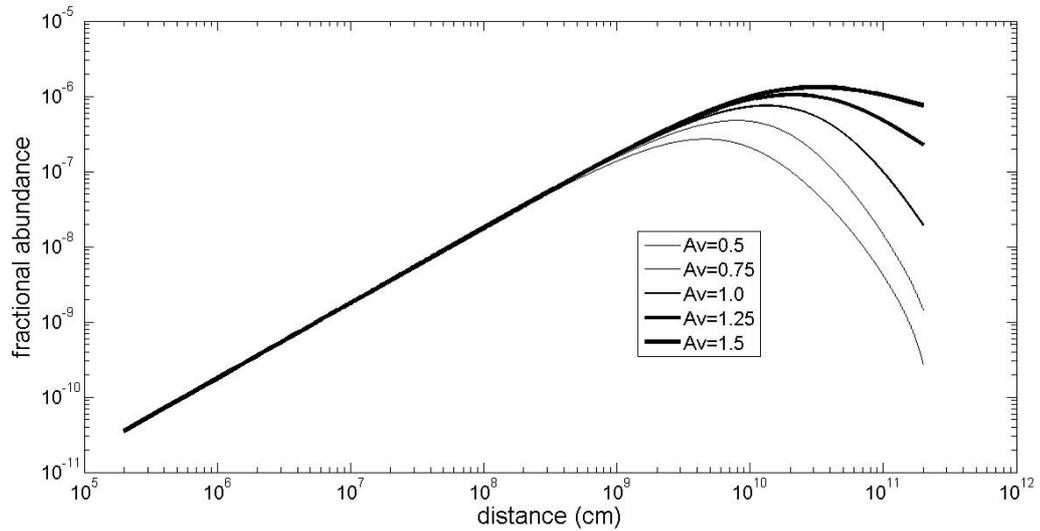

Figure 5. OH$^+$ fractional abundance profiles for different $A_v$ values. With C=$10^9$ and C/O = 1.0

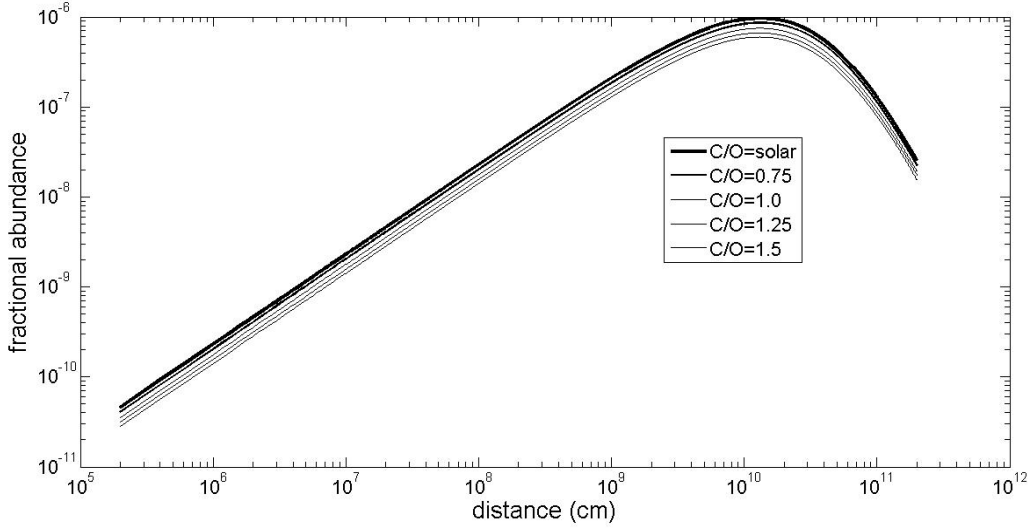

Figure 6. OH$^+$ fractional abundance profiles for different C/O ratios, with C=$10^9$ and $A_v$ = 1.0

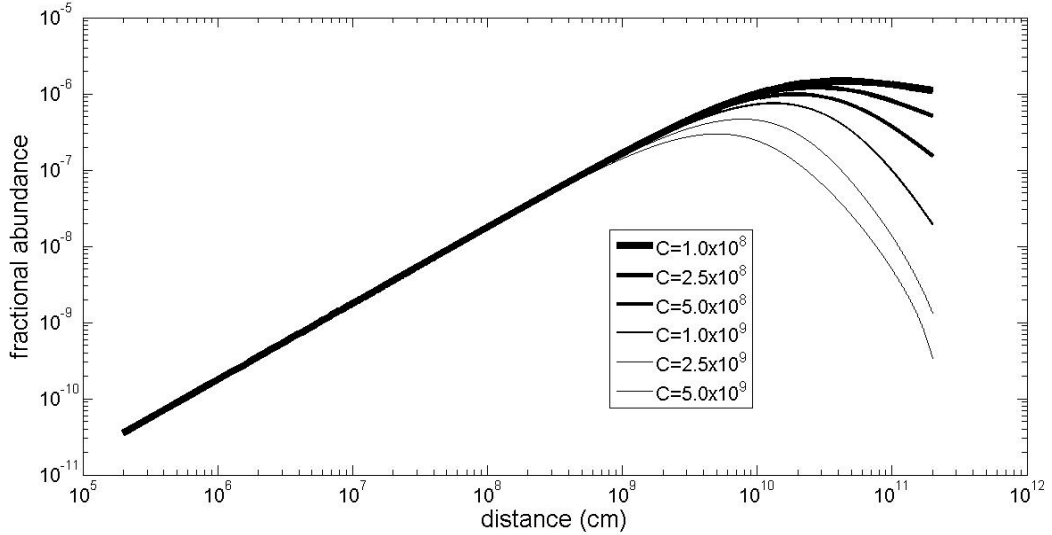

Figure 7. OH$^+$ fractional abundance profiles for different correction factors (*C*), with C/O=1.0 and $A_v$ = 1.0

For other Hot Jupiters with less intense XUV radiation field, the total column density could be even higher, provided that the effect of decreased photodestruction rate compensates the decreased mass loss rate. But, on the other hand, low density Hot Jupiters, would present high mass loss rates; a simple mass loss estimate by Sanz-Forcada et al. (2011), indicates that the mass loss is inversely proportional to the planet's bulk density (ρ), as represented by Eq. (5):

$$\dot{M} = \frac{3 F_{XUV}}{4 G \rho} \tag{5}$$

where G is the gravitational constant, and $F_{XUV}$ is incident XUV flux. And very low density Hot Jupiters and Hot Saturns, approaching 1/10 of Jupiter's density and lower are an observational fact (Anderson et al., 2010, Latham et al., 2010, Mandushev et al., 2007). Finally, H$_2$ survival in the exosphere is facilitated by less intense XUV radiation field.

## 4. Conclusions

The results of our simulations for chemical and photochemical reactions along the atmospheric mass loss of HD 209458b suggest that many species, especially small ions with oxygen, are formed with high abundances in a region where the temperature is low enough and the number density is still high enough for reactions to proceed. The formed molecules are then rapidly dissociated by the UV flux from the host star, with formation and destruction encapsulated in a radial distance of less than $10^7$ km. This conclusion is conditioned to the presence of residual molecular hydrogen along the mass loss, a premise that is supported by some observational evidence, but not yet proven correct. The species which reaches the maximum abundance, $OH^+$, is expected to exhibit column densities of ~ $10^7$ $cm^{-2}$, or perhaps even higher if the plasma dilution is attenuated by a magnetosphere, which would enable detection by current instruments, or if enough dust form the lower atmosphere survives the travel to the exosphere. In the event of the detection of $OH^+$ in the mass loss of Hot Jupiters, the models of the chemistry of the outer atmospheres of these planets would have to be revised. Higher values of the total column density of $OH^+$ and other species could be expected for low density Hot Jupiters subject to less intense XUV radiation.

The results for benzene production through the ion-molecule mechanism indicate that formation of carbon chain molecules is not significant in this environment, even with high values of C/O ratios. Since benzene is the basis for PAH production, we conclude that PAH formation is probably also inhibited.

## Acknowledgments

Vincent Bourrier, Isabel Aleman, Gelvam André Hartmann.